# JCAVE: A 3D INTERACTIVE GAME TO ASSIST HOME PHYSIOTHERAPY REHABILITATION

Lamiaa A. Elrefaei[1, 2], Bshaer Azan[1], Sameera Hakami[1], Safiah Melebari[1]

[1]Computer Science Department, Faculty of Computing and Information Technology, King Abdulaziz University, Jeddah, Saudi Arabia
[2]Electrical Engineering Department, Faculty of Engineering at Shoubra, Benha University, Cairo, Egypt

***ABSTRACT***

*The purpose of this paper is to investigate the applicability of applying gamification approach on the physiotherapy rehabilitation. A new developing game called JCave was designed and developed for the prove of concept. The propose game target the children from six to twelve years of age who need physical therapy in their upper limbs. JCave is a finite and multilevel single-player 3D video game. The player's role is to collect jewels from a cave and increase his/her score by performing physical therapy exercises. The game uses Xbox360 Kinect as a motion capture camera to observe gestures and track the child. Automatic gesture recognition algorithms are implemented for elbow flexion-extension exercises and shoulder flexion, which are the active range of motion (AROM) exercises for both the right and left arms. The JCave game is implemented using Unity3D and Blender to design 3D model objects.*

## 1. INTRODUCTION

The use of new technology such as game produces a significant improvement in healthcare. Games are used around the world, and new ideas and technical information are discovered every year; thus, games have a variety of definitions and genres [1], [2]. The oldest definition is from 1930. However, all of the definitions agree that a game is a set of rules and includes one or more players. One definition is as follows: "An interactive structure of endogenous meaning that requires players to struggle toward a goal" [1]. Moreover, the definition of a video game is the following: "X is a videogame if it is an artifact in a visual digital medium, is intended as an object of entertainment, and is intended to provide such entertainment through the employment of one or both of the following modes of engagement: rules and objective gameplay or interactive fiction" [2].

Game genres based on [1] are divided into three sections. The first section is general, in which all entertainment, educational, experimental, research, and operational types of games belong. The second section is players, in which a game is classified as single or multi-player. The third section depends on the game goals and the conditions: is it finite or infinite. Game genres, as in [2], are divided into several types: interactive faction, role-playing, adventure, real-time strategy, and algorithmic games.

The term 'Serious' applied to games denotes the application of gaming technologies and playful design strategies to domains of society and culture that are traditionally not associated with entertainment such as the medical domain [3]. This paper presents the design and implementation of a 3D interactive game called JCave using Xbox Kinect as a moving capture device. JCave's target users are children from six to twelve years of age who need physical therapy in an upper limb. The aim of JCave is to motivate the children to do their exercises at home in an easy, fun and exciting way. Based on [1] and [2], our serious JCave game is classified as a single-player, finite, adventure, and real-time game.

## 1.1. Motivation

Children generally like to discover new things, and they love adventure. When they are playing, they may fall, and the natural reflex is for them to throw their hands out in an attempt to stop the fall. Therefore, they are at risk for fracturing their arms [4]. Fractures of the upper extremities are much more common than those in the lower extremities in children. In the arm, elbow fractures are the most common [4], [5].

In some situations, after surgery or after removing a cast, the physiotherapist may prescribe extra exercises at home. However, children often become bored. Moreover, getting them to do the exercises is difficult. Therefore, rehabilitation takes a long time. To resolve this problem, interactive video games are used to recover from injuries and complete physiotherapy in a safe and efficient way [6].

Physiotherapy (physical therapy) is a branch of rehabilitation. It is the science of treating people with physical special needs due to illness, injury, or surgery and helps them maximize function, improve mobility, and relieve pain. Conditions or health problems that affect mobility, function, or cause disability may require physical therapy, such as back and neck injuries, fractures, arthritis, burns, and stroke [7]. The benefits of physiotherapy include building strength, improving balance, maintaining flexibility, and maximizing independence [8]. Some physiotherapy types do not require the use of special equipment or involvement from a physical therapist. Therefore, patients can perform their therapy at home. Range of motion (ROM) exercises are examples of at-home physical therapy [9], [10].

The ROM is the distance that the joint can be moved in a certain direction. It measures how far a patient can move his/her joints. These exercises help maintain flexible joints, reduce pain, and improve balance and strength. Before the treatment begins, the physical therapist generally measures the patient's range of motion. There are three types of ROM exercises [10]:

1. Passive ROM: The physical therapist moves the patient's joint and no active movement comes from the patient.
2. Active-assistive ROM: Some assistance from the physical therapist is provided. The patient can move his or her limb but cannot complete the full range of motion because of weakness or pain.
3. Active ROM (AROM): All movements are performed by the patient.

The physical therapist may continually provide verbal cues on how to perform the exercises properly [9], [10]. Most injuries that children incur are in the upper extremities (shoulder, elbow, and wrist) because children are typically active individuals. The treatment depends on the type of fracture and the degree of displacement. The doctor may immobilize the arm in a cast or perform surgery. After surgery and removing the cast, range of motion exercises are recommended to improve the range of motion in the joint [11]. Table 1 illustrates the upper limb joints and their AROM exercises.

In this paper, a 3D interactive game called JCave is designed using Xbox Kinect as a moving capture device. It is interactive single-player adventure, and the end is finite. It focuses on visual objects, and the goal is to increase the player's score to advance to the next level. JCave helps decrease the duration of the patient's physical therapy rehabilitation program. JCave's target users are children from six to twelve years of age who need physical therapy in an upper limb. Elbow flexion and extension and shoulder flexion AROM exercises are supported in the JCave game. The aim of JCave is to motivate the children to do their exercises at home in a fun and exciting way. The player's role in JCave is to collect jewels from a cave to increase his/her score and complete all levels.

This paper presents and explains the game development of the proposed JCave game focusing on the AROM exercises analysis and automatic recognition and the gamified environment.

Table 1. AROM exercises for the upper limb joints

| Upper limb joint | AROM exercises |
|---|---|
| Elbow | *Flexion and *extension exercise |
| Shoulder | Abduction and adduction exercise, *flexion exercise, and extension exercise |
| Wrist | Flexion and extension exercise |

* Exercises supported in the JCave game

### 1.2. Paper Structure

The rest of this paper is organized as follows: Section 2 presents the background and related work. Then, the concept and description of JCave game is presented in Section 3. Section 4 details the AROM exercises gesture analysis and automatic recognition. Section 5 explains the configuration/customization and physical therapy environments of JCave game. The Experiments and results are presented in Section 6. Finally, Section 7 concludes the paper and highlights the future work.

## 2. Background And Related Work

### 2.1. Limb Motion Capturing

There are two major approaches for motion capturing:

1. Contact based devices: are based on the physical interaction of the user with the interfacing device i.e. glove, accelerometers, multi-touch screen, Nintendo Wii- Remote [12].
2. Vision based devices: using devices relies on captured video sequence by one or several cameras for interpreting and analysing the motion i.e. motion capture camera Kinect and PrimeSense 3D camera [12].

Both approaches have their advantages and disadvantages. The main disadvantage of contact-based devices is the health hazards which are caused by its devices like mechanical sensor material which raise symptoms of allergy, magnetic devices which raise risk of cancer etc. On the other hand, vision-based devices though are user friendly but suffer from configuration complexity and occlusion problems, hence more privileged for usage in long run [12].

### 2.1.1. Motion Capture Cameras

There are many sensors that track people, e.g., ZED [13] (Fig. 1(a)), Swiss Ranger [14], [15] (Fig. 1(b)), Wiimote [16] (Fig. 1(c)), OptiTrack [17] (Fig. 1(d)), and Kinect [18] (Fig. 1(e)). These sensors are types of 3D motion capture cameras. Table 2 compares these sensors in terms of the minimum and maximum distance between the person and the sensor, the number of frames the sensor captures per second, the environment that the sensor should be in, if a marker is needed or not, resolution, the degree of viewing range, and cost.

Based on the comparison in Table 2, Wiimote and OptiTrack are not appropriate sensors for the JCave game because the target user has injuries to an upper limb, making it difficult for them to hold the marker. The Kinect platform is compatible with JCave because it is familiar to the users, is inexpensive, and does not need a marker.

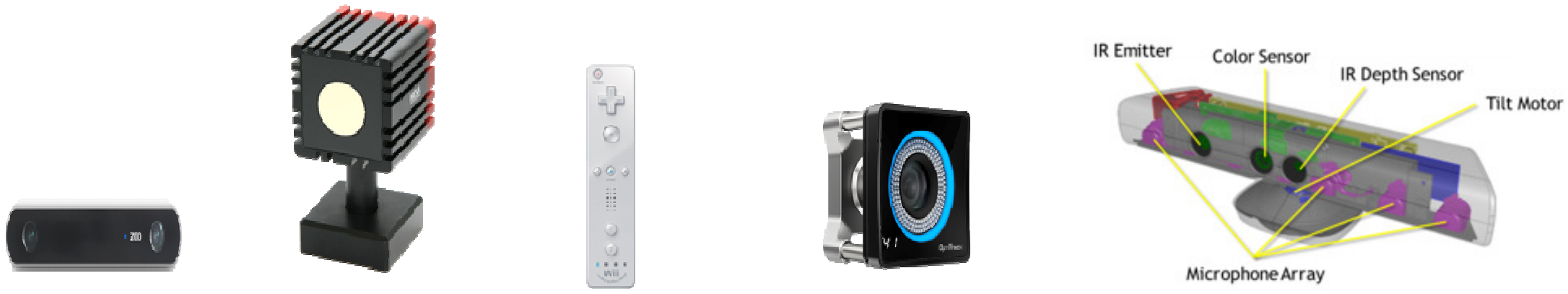

(a) ZED camera | (b) Swiss Ranger camera | (c) Wiimote | (d) OptiTrack camera | (e) Kinect and its Components

Figure 1. Motion capture sensors

### 2.1.2. Kinect sensor

The Kinect sensor is a type of input device. It is a 3D motion capture camera that was developed by Microsoft. It was released in November 2010 with Xbox and was first designed for natural interaction in a video game environment [19]. In 2012, the Kinect sensor for Microsoft Windows was released. Fig. 1(e) shows the four main components of the Kinect sensor: depth sensor, RGB camera, tilt motor, and multi-array microphone [18], [20], [8], [21].

Table 2. A comparison of the motion capture sensors shown in Figure 3

| | Kinect | ZED | Swiss Ranger | Wiimote | OptiTrack |
|---|---|---|---|---|---|
| Distance | 0.8 – 4 m | 0.07 – 20 m | 7.5 m | Up to 5 m | 20 m |
| Frames per second | 30 fps | 15-100 fps | 30-100 fps | - | 100 fps |
| Environment | Indoor | Indoor and outdoor | - | Indoor | Medical center |
| Marker | No | No | No | Yes | Yes |
| Resolution | 640x480 pixels | 1344x376 - 4416x1242 pixels | 124x160 pixels | - | 640x480 pixels |
| Viewing range | 57° horizontally and 43° vertically | 110° | 43° horizontally and 46* vertically | - | 360° |
| Cost | $110 | $440 | - | $300 | $600 |

## 2.2. Related Work

There are many rehabilitation systems in the literature that employ limb motion capturing. In [3], the authors proposed an application that uses a single custom-made sensor node consisting of an Inertial Measurement Unit (IMU) attached on a lower limb in order to capture its orientation in space in real-time and provides input to a game on an Android mobile device. Their aim is to increase patient engagement during physiotherapy by motivating the user to participate in a game. In [22], the authors proposed an Augmented Reality (AR) system for the rehabilitation of hand movements which have been impaired due to illness or accident. Their system uses a standard computer, a webcam and two wireless 5DT data gloves. their system requires specialized technical support and maintenance. In [23], a high-cost motion capture suit requiring technical set-up was used for representing changes in human motion symmetry during injury rehabilitation. The main drawback of these systems is the health hazards resulted from the contact-based devices [12]. To overcome this drawback our JCave game is using Xbox Kinect as a motion capture device which relies on captured video sequence analysis without direct contact.

Also, there are many games for physical therapy rehabilitation using camera sensors, and JCave was inspired by them. *ReaKing* [24] is an interactive game using Microsoft Kinect V2 to track people. It is designed to help elderly people perform physical activity. All movements of the patient are recorded so that the physiotherapist can check the progress of the patient. *Mira* [25] is a system for physiotherapy that uses interactive games based on the best clinical practices and the expertise of physiotherapists. Using Microsoft Kinect for Xbox360, the patients are able to progress to different levels within the games by performing rehabilitating movements. The *Therapy store* [26] is a therapy system that uses Xbox360 Kinect to monitor a patient's joints at home. It was designed for patients with cerebral palsy and those receiving rehabilitation after a long period of bedrest.

The differences between these games and the proposed game, JCave, is that JCave is a home-based game, which means that there is no need to go to a physiotherapy centre to do the exercises. Moreover, JCave takes into consideration the target users and their conditions by not requiring the player to use his/her movements to pre-set the game. JCave supports a mouse to pre-set the game.

## 3. JCAVE GAME CONCEPT AND DESCRIPTION

To choose a suitable concept and idea for the game, we should consider who the target users are and their preferences. The game should be preferred by children from six to twelve years of age and suitable for physiotherapy. The game should not be too exiting to make the child move violently. Moreover, the game should have levels to meet the need of the physiotherapy program and encourage the child to continue doing the exercises.

JCave is a 3D video game for a single player. The player's role is to collect jewels from a cave to increase his score and complete all levels. The player must reach a specific score to finish the level and collects the jewels using a hook that moves automatically on a pendulum from which the hook extends to catch the jewel. A simple description of the JCave game is shown in Figure 2 with the following aspects:

- JCave is a 3D interactive game for children who have an upper limb fracture.
- Elbow flexion and extension and shoulder flexion exercises are chosen as the physiotherapy exercises because the child can perform the exercises alone at home.
- A motion capture camera is used to capture the gestures and track the child. We use Xbox360 Kinect because it is popular, and most children have access to it. The adapter to connect to a PC is the only required addition.
- Microsoft Kinect SDK 1.8 is used as the driver for the Windows operating system.
- A gesture automatic recognition algorithm for the AROM exercises is implemented using the skeleton data of the player provided by Xbox360 Kinect.

- We design and implement a real-time, interactive, single-player, multilevel game using Unity3D and visual studio programs with the C# language.
- We implement the JCave game for a PC with Windows 8 or later, a 64-bit processor, at least 4 GB of memory, a dual-core 3.1 GB or faster processor, and a DX11-capable graphics adapter.

## 4 AROM EXERCISES GESTURE ANALYSIS AND AUTOMATIC RECOGNITION

### 4.1. Exercises Gesture Analysis

We analyze the elbow flexion and extension and shoulder flexion exercise gestures using the Xbox360 Kinect motion capture camera to capture the gestures and track the child.

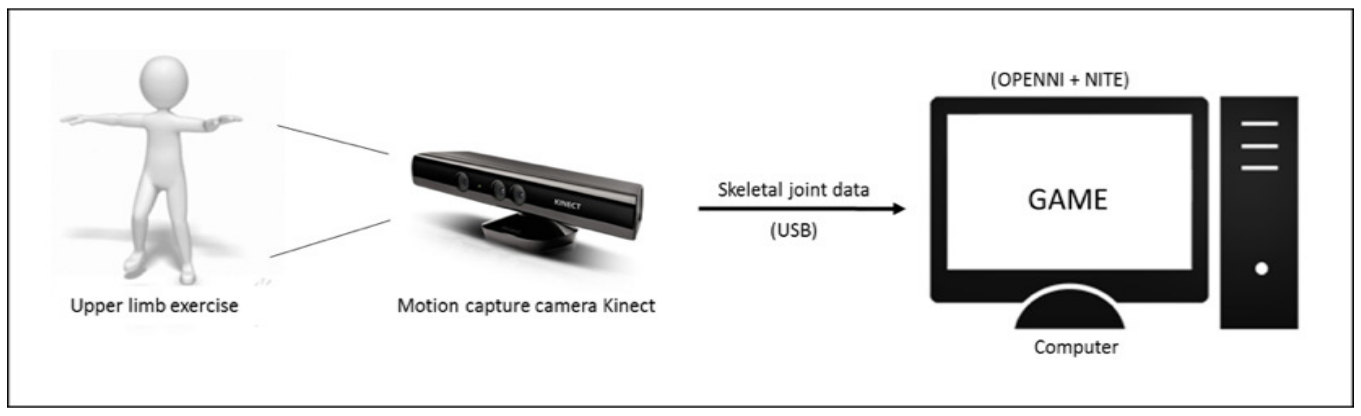

Figure 2. Simple description of the JCave game

Kinect supports three spaces: color space, depth space, and skeleton space. A Kinect streams color, depth, and skeleton data one frame at a time. In JCave, the skeleton space is used to represent the ROM exercise using skeleton data. The skeleton data contain 3D positions of human skeletons. The position of a skeleton and each of the skeleton joints (if active tracking is enabled) are stored as (x, y, z) coordinates [27].

The skeleton space, as shown in Figure 3(a), is a right-handed coordinate system that places the Kinect sensor at the origin. The positive z-axis extends in the direction in which the Kinect sensor is pointed. The positive y-axis extends upward, and the positive x-axis extends to the left. Figure 3(b) shows the Kinect skeleton space, where the origin (0,0,0) is the location of the IR depth sensor [27].

In the JCave game, we implement two AROM exercises: elbow flexion/extension and shoulder flexion exercises. To analyze the AROM exercise movement based on the Kinect skeleton space, we consider the Cartesian coordinate system of the user as a mirror of the Kinect skeleton space, as shown in Figure 3(c).

A tracked skeleton provides detailed information about the position of twenty joints of the user's body, as shown in Figure 3(d) [28]. The skeleton data for any joint consist of the positions of the joint in the 3D coordinate system.

In the following analysis, we use the skeleton joint position symbols Hand.X, Hand.Y, and Hand.Z to denote the position of the hand on the X-axis, Y-axis, and Z-axis, respectively. Similar symbols related to the elbow and shoulder positions are also used: Elbow.X, Elbow.Y, Elbow.Z, Shoulder.X, Shoulder.Y, and Shoulder.Z. By defining the exercise movement with the position of the hand, elbow, and shoulder in the Cartesian coordinate system, as shown in Figure 3(c), the game uses the Kinect skeleton data for the user and verifies if the exercise movement is correct.

### 4.1.1. Elbow Flexion/Extension exercise Analysis

The elbow flexion/extension exercise begins with the arm relaxed next to the body. The arm is gently moved until the forearm makes contact with the muscles of the arm. Then, the arm is extended to be next to the body again, as shown in Figure 4(a). The exercise is divided into two parts: flexion and extension. Flexion refers to a movement that decreases the angle between two body parts. The flexion at the elbow decreases the angle between the ulna and the humerus, as shown in Figure 4(b). Extension refers to a movement that increases the angle between two body parts. The extension at the elbow increases the angle between the ulna and the humerus, as shown in Figure 4(b).

Both the flexion and extension parts are further divided into two segments: hand below elbow (hand down) and hand above elbow (hand up), as shown in Figure 4(b).

The exercise is described in all directions to place limits on all arm joint movements by observing the relationships between the positions of the joints (hand, elbow, and shoulder) along the X, Y, and Z directions.

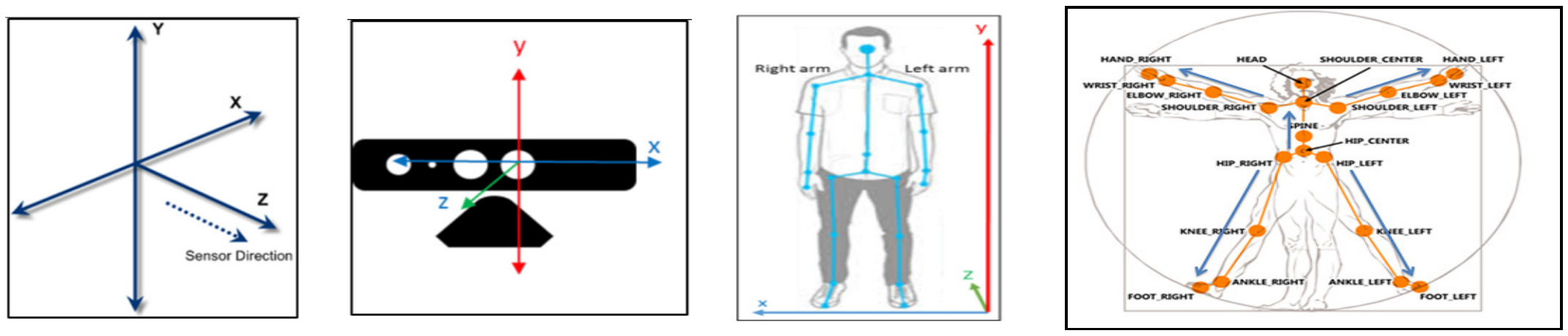

(a) Skeleton space [29] (b) Kinect skeleton space (c) User space (d) Skeletal joints [29]

Figure 3. Kinect skeleton

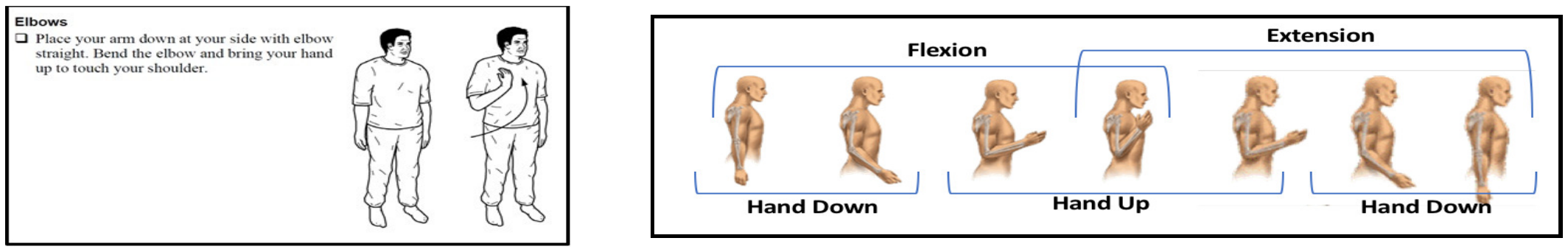

(a) (b)

Figure 4. (a) Elbow flexion/extension exercise (b) Elbow flexion/extension segments

#### 4.1.1.1. X-axis Analysis

#### Hand below elbow (hand down) position along the X-axis

The arm does not typically form a straight line from the shoulder to the hand; it forms a carrying angle (A). The carrying angle of the elbow, as shown in Figure 5(a), is the angle made by the axes of the arm and the forearm when the elbow is in full extension. The carrying angle is typically greater in females than males. When the arms are held out at the sides of the body, the hands should, on average, be approximately 5 to 15 degrees away from the body. This is the normal carrying angle of the elbow. Considering that the angle $(A)$ is 15 degrees, the maximum angle the child should have to perform the exercise is $A + C$, as shown in Figure 5(a).

As shown in Figure 5(a), the maximum distance (*M*) of the hand along the X-axis when the hand is down is calculated as follows:

$$M=|X3-Elbow.X|=|Elbow.Y-Y3|*\tan(A+C) \quad (1)$$

where angle *C* is calculated as follows:

$$C=B=\tan^{-1}(|Shoulder.X-Elbow.X|/|Shoulder.Y-Elbow.Y| \quad (2)$$

Then, the condition for the right arm is Elbow.X $\leq$ Hand.X $\leq$ Elbow.X + M; for the left arm, the condition is Elbow.X - M $\leq$ Hand.X $\leq$ Elbow.X.

### Hand above elbow (hand up) position along the X-axis

The same *M* distance calculation is applied for the hand up position along the X-axis, although it is used in the opposite direction, as shown in Figure 5(b). The condition for the right arm is Elbow.X - M $\leq$ Hand.X $\leq$ Elbow.X, and the condition for the left arm is Elbow.X $\leq$ Hand.X $\leq$ Elbow.X + M.

### 4.1.1.2. Y-axis Analysis

### Hand below elbow (hand down) position along the Y-axis

As shown in Figure 5(a), the condition for both arms is Hand.Y < Elbow.Y.

### Hand above elbow (hand up) position along the Y-axis

According to Figure 5(b), when the hand is above the elbow from the positive to negative Y direction, the order of the body parts is shoulder, hand, and elbow. The shoulder and hand may have the same position, or the hand may be above the shoulder. To cover all possibilities and prevent an incorrect movement, we consider the hand to be above the shoulder with a constant k. The suitable heuristic value of k after trying values from 0 to 1 is 0.2. Thus, the condition for both arms is Hand.Y $\leq$ Shoulder.Y + k.

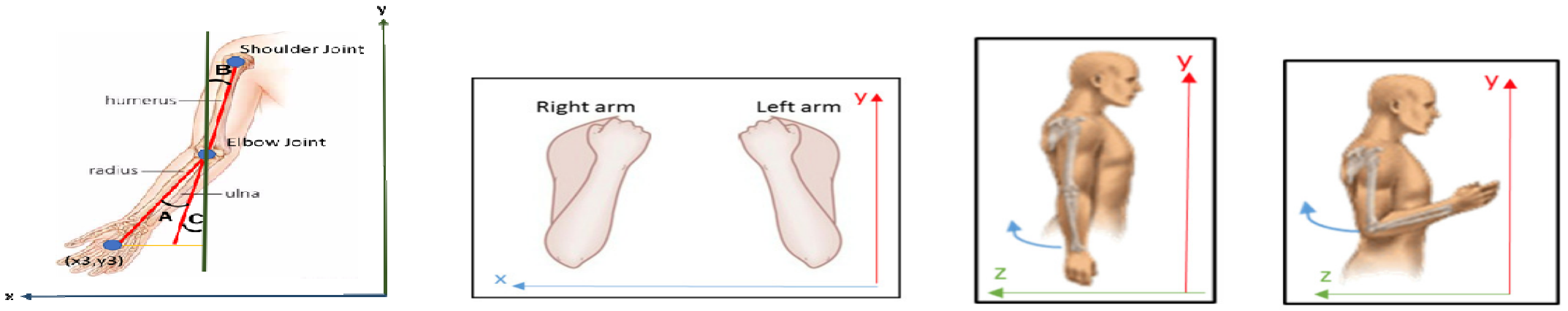

(a) Carrying angle and arm joints in the XY plane

(b) Hand above elbow positions in the XY plane

(c) Hand and elbow positions in the YZ plane

(d) Elbow and shoulder positions in the YZ plane

Figure 5. Elbow flexion/extension exercise analysis

### 4.1.1.3. Z-axis Analysis

### Hand below elbow (hand down) position along the Z-axis

As shown in Figure 5(c), the condition for both arms is Hand.Z $\leq$ Elbow.Z.

## Hand above elbow (hand up) position along the Z-axis

As shown in Figure 5(d), the condition for both arms is Elbow.Z ≤ Shoulder.Z.

A summary of the elbow flexion/extension exercise analysis is presented in Table 3. A complete exercise is performed with three segments: hand below elbow (hand down), hand above elbow (hand up), hand below elbow (hand down), as illustrated in Figure 4(b). The exercise is considered correct if the three segments are performed continually without interruption or incorrect movements.

### 4.1.2. Shoulder Flexion Exercise Analysis

The shoulder flexion exercise begins with the arm relaxed next to the body. The elbow is kept straight, and the arm is raised up in front of the body and toward the ceiling. The arm is gently stretched upward and then brought down, as shown in Figure 6(a). The exercise is divided into three segments: hand below shoulder (hand down), hand above shoulder (hand up), and hand below shoulder again (hand down), as shown in Figure 6(b).

Again, the exercise is described in all directions to place limits on all arm joint movements by observing the relationships between the positions of the joints (hand, elbow, and shoulder) along the X, Y and Z directions.

Table 3. A summary of the elbow flexion/extension exercise analysis

| Hand position | Analysis direction | Involved joints | Rules for correct joint positions during movements | |
|---|---|---|---|---|
| | | | Right arm | Left arm |
| Hand below elbow (hand down) | X-axis | Hand and elbow joints | Elbow.X ≤ Hand.X ≤ Elbow.X + M | Elbow.X - M ≤ Hand.X ≤ Elbow.X |
| | Y-axis | Hand and elbow joints | Hand.Y < Elbow.Y | Hand.Y < Elbow.Y |
| | Z-axis | Hand and elbow joints | Hand.Z ≤ Elbow.Z. | Hand.Z ≤ Elbow.Z. |
| | | | | |
| Hand above elbow (hand up) | X-axis | Hand and elbow joints | Elbow.X - M ≤ Hand.X ≤ Elbow.X | Elbow.X ≤ Hand.X ≤ Elbow.X + M |
| | Y-axis | Hand and shoulder joints | Hand.Y ≤ Shoulder.Y + k | Hand.Y ≤ Shoulder.Y + k |
| | Z-axis | Elbow and shoulder joints | Elbow.Z ≤ Shoulder.Z | Elbow.Z ≤ Shoulder.Z |

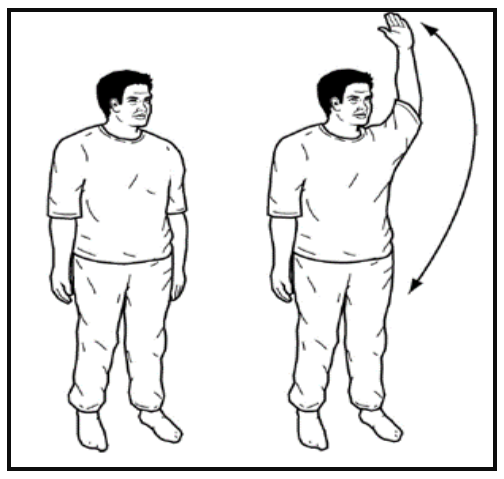

(a)

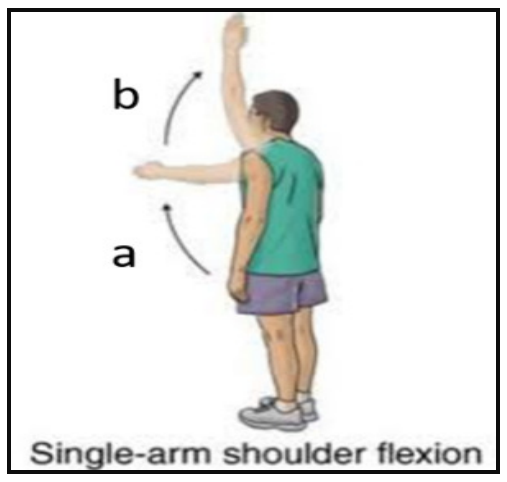

(b)

Figure 6. (a) Shoulder flexion exercise

(b) Shoulder flexion segments (a. hand Down, b. hand Up)

#### 4.1.2.1. X-axis Analysis

When the right hand is above the shoulder or is below the shoulder, the hand, elbow, and shoulder have the same X position or are ordered from highest to lowest X position as follows: hand, elbow, and shoulder.

When the left hand is above the shoulder or is below the shoulder, the hand, elbow and shoulder have the same X position or are ordered from highest to lowest X position as follows: shoulder, elbow, and hand.

The conditions for the right arm are Hand.X ≥ Elbow.X, Hand.X ≥ Shoulder.X, and Elbow.X ≥ Shoulder.X. The conditions for the left arm are Hand.X ≤ Elbow.X, Hand.X ≤ Shoulder.X, and Elbow.X ≤ Shoulder.X.

#### 4.1.2.2. Y-axis Analysis

When the right or left hand is below the shoulder, the hand, elbow, and shoulder are ordered from highest to lowest Y position as follows: shoulder, elbow, and hand.

When the right or left hand is above the shoulder, the hand, elbow, and shoulder are ordered from highest to lowest Y position as follows: hand, elbow, and shoulder.

The conditions for both arms in the hand down position are Hand.Y ≤ Elbow.Y, Hand.Y ≤ Shoulder.Y, and Elbow.Y ≤ Shoulder.Y. The conditions for both arms in the hand up position are Hand.Y > Elbow.Y, Hand.Y > Shoulder.Y, and Elbow.Y > Shoulder.Y.

#### 4.1.2.3. Z-axis Analysis

When the right or left hand is above the shoulder or is below the shoulder, the Z position of the hand is greater than or equal to the Z position of the elbow. The condition is Hand.Z ≥ Elbow.Z. The complete production rules are summarized in Table 4.

Table 4. A complete analysis for the shoulder flexion exercise

| Hand position | Analysis direction | Involved joints | Rules for correct joint positions during movements | |
|---|---|---|---|---|
| | | | Right arm | Left arm |
| Hand below shoulder (hand down) | X-axis | Hand, elbow, and shoulder joints | Hand.X ≥ Elbow.X and Hand.X ≥ Shoulder.X and Elbow.X ≥ Shoulder.X | Hand.X ≤ Elbow.X and Hand.X ≤ Shoulder.X and Elbow.X ≤ Shoulder.X |
| | Y-axis | Hand, elbow, and shoulder joints | Hand.Y ≤ Elbow.Y and Hand.Y ≤ Shoulder.Y and Elbow.Y ≤ Shoulder.Y | Hand.Y ≤ Elbow.Y and Hand.Y ≤ Shoulder.Y and Elbow.Y ≤ Shoulder.Y |
| | Z-axis | Hand and elbow joints | Hand.Z ≥ Elbow.Z | Hand.Z ≥ Elbow.Z |
| | | | | |
| Hand above shoulder (hand up) | X-axis | Hand, elbow, and shoulder joints | Hand.X ≥ Elbow.X and Hand.X ≥ Shoulder.X and Elbow.X ≥ Shoulder.X | Hand.X ≤ Elbow.X and Hand.X ≤ Shoulder.X and Elbow.X ≤ Shoulder.X |
| | Y-axis | Hand, elbow, and shoulder joints | Hand.Y > Elbow.Y and Hand.Y > Shoulder.Y and Elbow.Y > Shoulder.Y | Hand.Y > Elbow.Y and Hand.Y > Shoulder.Y and Elbow.Y > Shoulder.Y |
| | Z-axis | Hand and elbow joints | Hand.Z ≥ Elbow.Z | Hand.Z ≥ Elbow.Z |

## 4.2 AROM Exercises Gesture Automatic Recognition

For gesture recognition, Xbox Kinect 360 and Kinect for Windows SDK v1.8 are used to provide skeleton data of the player, and the C# programming language is used for implementation. Figure 7(a) presents the Kinect flowchart:

- First, the Kinect is checked, and the skeleton stream is enabled when the Kinect is connected.
- Then, the skeleton frame is opened, and the user is detected.
  - If the chosen exercise is the elbow flexion extension, the carrying angle A and the maximum distant M are calculated, using equations (2) and (1) respectively, to recognize the gesture.
  - If the chosen exercise is the shoulder flexion, the gesture recognition is initiated.

Based on the analysis for the two exercises, as presented in Sections 4.1.1 and 4.1.2, the algorithms for each segment, using the derived production rules listed in Tables 4 and 5, are written. For example, algorithm 1 lists the steps for the hand above elbow (hand up) segment for the right arm.

The gesture recognition flowchart is shown in Figure 7(b), where a complete exercise is performed in three segments: hand down, hand up, and hand down again. A maximum number of 100 frames (windowSize) are allowed to complete each segment. A complete exercise gesture is recognized only if the three segments are succeeded. The C# open-source code of the wave gesture from [29] is adopted and modified for the JCave gesture recognition, as shown in Figure 7(b). A screen shot of the implementation output of a complete elbow flexion/extension gesture recognition is shown in Figure 8(a) and that for a complete shoulder flexion gesture recognition is shown in Figure 8(b).

# 5 JCAVE GAME ENVIRONMENT

Three software tools, Unity3D, Blender, and Visual Studio2016, are used to implement the game:

- Visual Studio: This platform is used for the gesture recognition implementation and the coding portion of the project.
- Unity3D: This platform is used to create some objects and implement the entire game.
- Blender: This tool is used to model and animate the 3D objects.

The final JCave game consists of two main sections: the configuration/customization section and the physical therapy section. JCave game flowchart is shown in Figure 9.

## 5.1. JCave Configuration/Customization Section

In the configuration/customization section, the child can create new profile and save the player information or select from previously saved profiles. The child also setup some aspects such as character selection, AROM exercise type selection, right/left arm selection, level/sublevel selection and edit the number of the exercises as advised by the physiotherapist. The player uses the mouse and keyboard in the configuration/customization section. JCave configuration/customization section is shown in Figure 10.

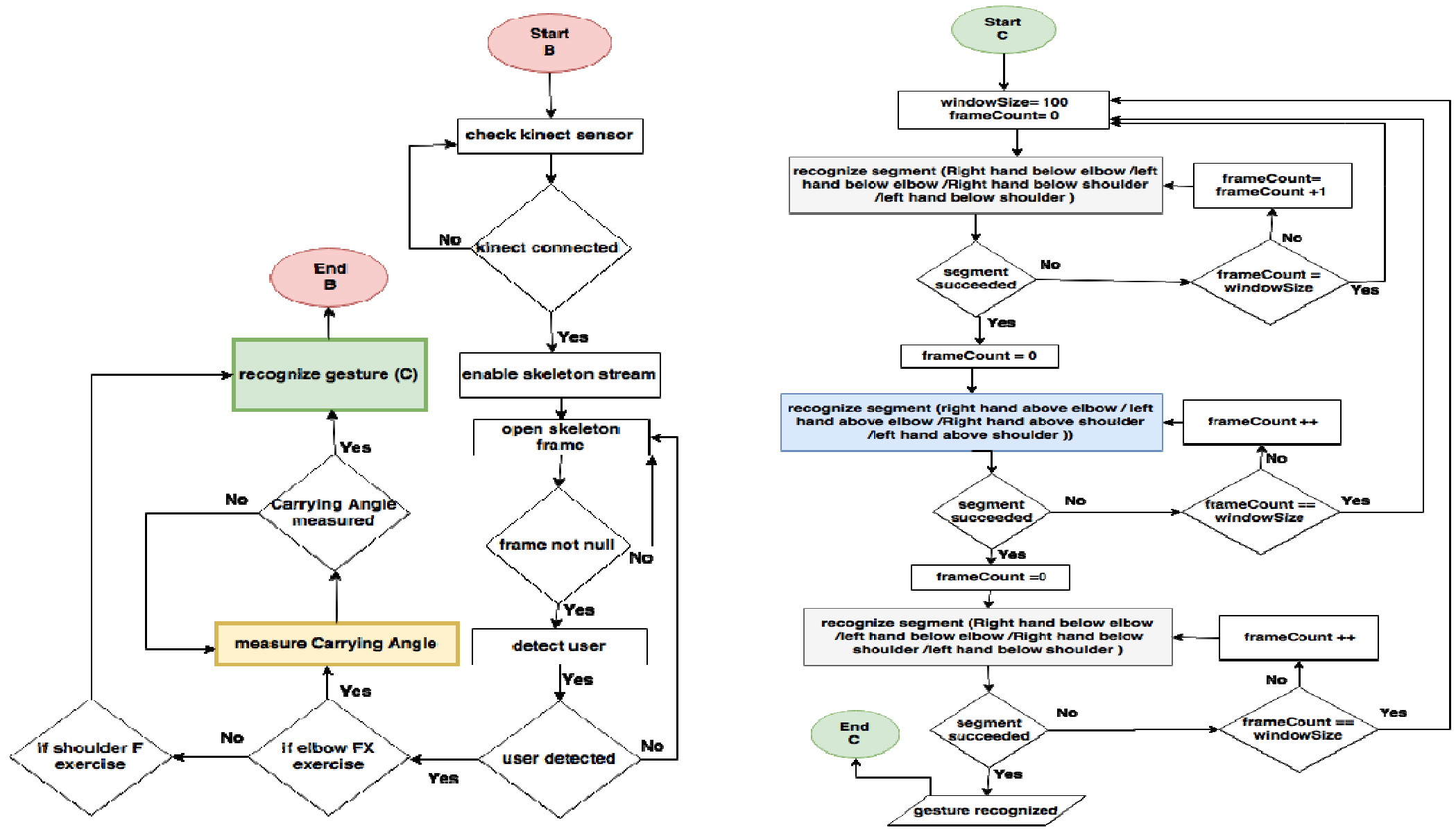

(a) (b)

Figure 7. (a) Kinect flowchart (node B) (b) Gesture recognition flowchart (node C)

---

**Algorithm 1: RightArm_HandAboveElbow_HandupSegment**

**Input:** skeleton, M //where skeleton is an array containing the skeleton joints, and M is the maximum distance the hand is extended along the X-axis.

**Output:** succeeded or failed //for correct or incorrect segments

---

```
if skeleton (position Y of right hand) > skeleton (position Y of right elbow) then
   if skeleton (position X of right hand) >= skeleton (position X of right elbow) – M && skeleton (position X of right hand) <= skeleton (position X of right elbow) then
      if skeleton (position Z of right elbow) <= skeleton (position Z of right shoulder) then
         if skeleton (position Y of right hand) <= skeleton (position Y of right shoulder) + 0.2 then
            return succeeded
return failed
```

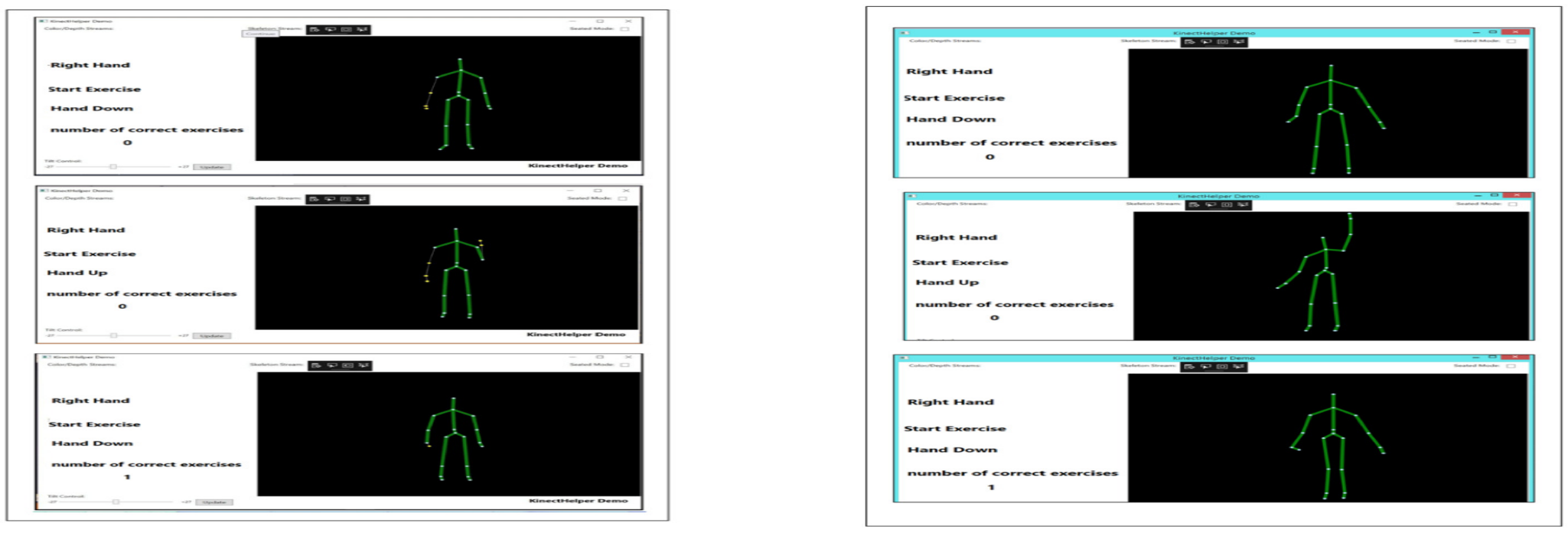

(a) A complete elbow flexion/extension gesture recognition

(b) A complete shoulder flexion gesture recognition

Figure 8. Screen shots of the implementation output of gesture recognition

### 5.2. Jcave Physical Therapy Section

For the physical therapy section, the main interface is shown in Figure 11 which contains the game environment. The JCave environment include the cave, jewels, character (who collects the jewels), the hook, and the gamification feedback elements (Figure 11). The hook moves automatically on a pendulum from which the hook extends to catch the jewel. The child performs the exercise and the Kinect sensor captures the player's movements. The real time skelton image from the Kinect is also displayed on the main interface, as shown in Figure 11, as a mean of self-visual feedback. If the exercise gesture is recognized correctly by the system, the hook moves downward to collect the jewel. The hook movement steps are illustrated in Algorithm 2. If the hook hits the jewel, the score will increase depending on the size and the type of the jewel. The hook movement for a complete correctly recognized elbow flexion/extension gesture is shown in Figure 12. If the player achieves the score of the sub-level and finishes the number of exercises required each day, he/she will win and proceed to the next sub-level (Figure 13(a) shows a visual hint from the system for winning). The player loses when he/she performs an exercise twice and does not reach the score required of the sub-level and the system shows a visual hint to try again (Figure 13(b)).

The JCave game has two levels. In each level, there are twelve sub-levels/stages. In level 1, The jewel has fixed locations in the environment and in level 2, the locations of the jewels change in the environment. The player has to complete the first level to begin the second level. To win, the player must perform at least N exercises and at most N*2 exercises and reach the sub-level's score, where N is the number that the player enters as the number of exercise repetitions when creating his/her profile. The player loses if he/she performs the exercise N*2 and does not reach the sub-level's score. The WinOrLose steps are illustrated in Algorithm 3.

To represent the collected jewel as a score, three sizes of six different jewels shown in Table 6 are created. Each size is associated with a different score value, as shown in Table 6. The collectedScore is used to calculate the player's score using (3):

$$collectedScore=collectedScore+(size+1)*10+(JewelIndex*10) \tag{3}$$

where the size is 0, 1, or 2, and JewelIndex is the index of the collected jewel, as shown in Table 5.

From Table 5, the player must collect a score of N*10 to win, which is the worst case if he/she collects a jewel of index 1 with size 0.

## 6 Experiments And Results

### 6.1 Arom Exercises Gesture Automatic Recognition Experiment

For the system to correctly recognize the exercise gestures, we must comply with the Kinect constraints [30]:

- The Kinect sensor should be on stable surface.
- The area between the player and the Kinect sensor should be clear.
- The player should stand approximately 1.4 m from the Kinect sensor.
- The room should be bright.
- The Kinect sensor should be placed between 0.6 m (2 feet) and 1.8 m (6 feet) from the floor.

- The sensor should not be placed in direct sunlight or within 0.3 meters of the audio speakers.

Further, the gesture recognition testing is applied according to the time needed to perform the exercise (t) in seconds, height of the testers (h) in centimeters, and the distance between the Kinect sensor and the testers (d) in meters. Table 6 shows the gesture recognition status. First, we set h = 148 cm and d = 1.5 m and change the time t. Second, we set d =1.5 m and change the height h at different times t. Finally, we set h = 148 cm and change d at different times t. The representativeness of the testers' heights in tests comes from the game scope and our target users (children from six to twelve years of age). It can be concluded from Table 6 that gesture recognition fails when the exercises are performed too slowly (>7 Sec) which encourage the child to perform the exercise within 3-6 sec. The gesture recognition also fails when the distance d equals to 3 m, but as the game is designed to be at-home (indoor environment), the distance d is guaranteed to be less than 3 m.

## 6.2 Usability Experiment

The purpose of this experiment is to measure the usability of the JCave game, to take notes, and provide additional improvements. The general goal is to ensure that lists and labels in the JCave game are easy to reach and navigate. The quantitative usability goals for the JCave game are to ensure that users are able to understand the meaning of the lists and labels and find the choices that they want in less than one minute with no more than two incorrect choices the first time they need it. They also remember the process correctly after the first time and select it in less than one minute. The general concern of the test is about understanding the JCave game interface and concept. The specific tasks of the test are: start the game, create new profile, edit profile, view help content, start playing, pause the game, resume the game, control the volume and music, exit play, delete profile, and exit from the game.

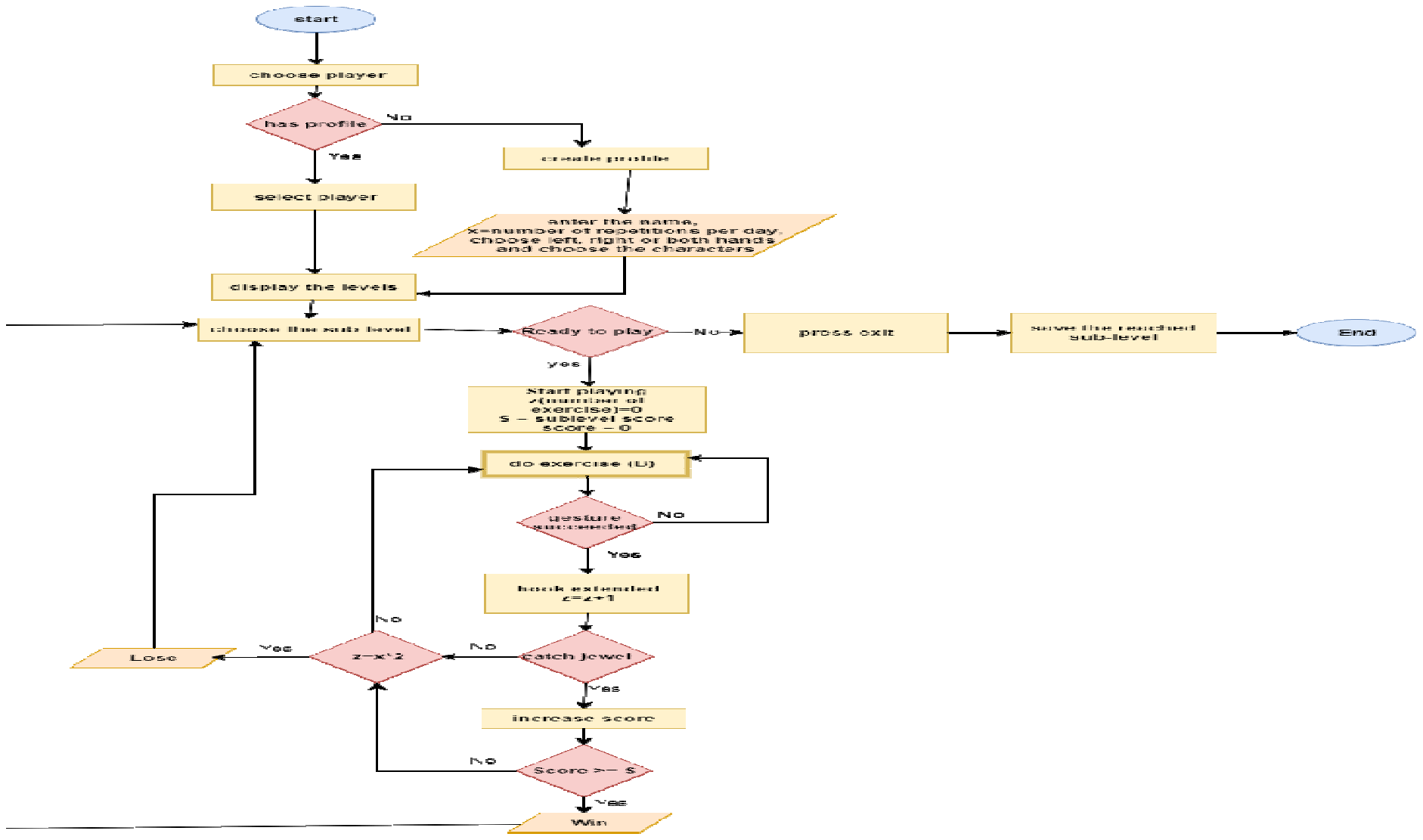

Figure 9. JCave game flowchart Node (B) is shown in Figure 7

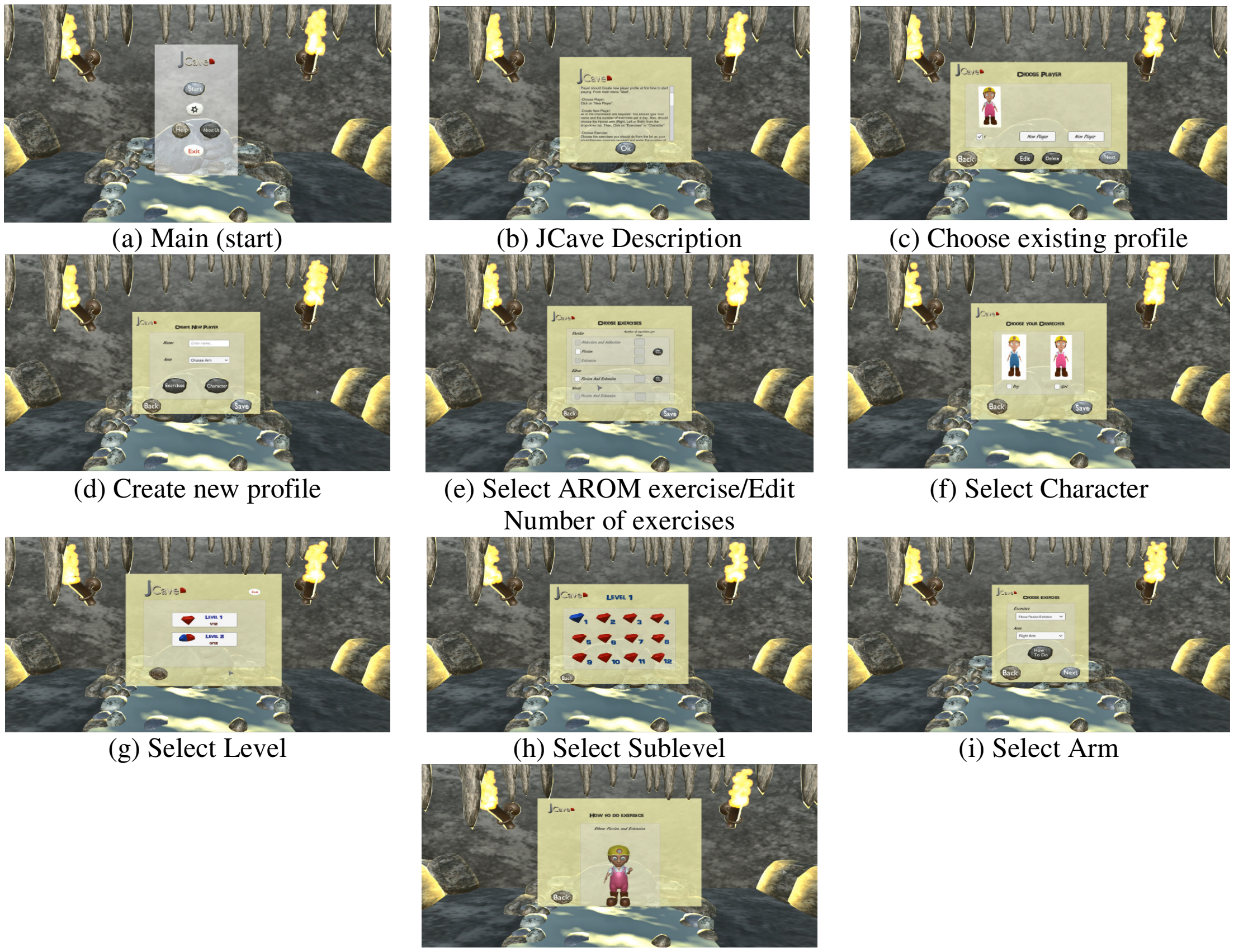

(a) Main (start) (b) JCave Description (c) Choose existing profile

(d) Create new profile (e) Select AROM exercise/Edit Number of exercises (f) Select Character

(g) Select Level (h) Select Sublevel (i) Select Arm

(j) How to do exercise

Figure 10. JCave configuration/customization section

---

**Algorithm 2:** HookMovement

**Input:** *done*, *triggerEvent*

//where *done* is a Boolean value (true if the player does the exercise), and *triggerEvent* is a static Boolean value (true if the hook hits the jewel)

**Output:** move the hook downward

```
count ← 0
extAmount ← 0.01
if done = true
  while count < 200 and triggerEnent = false do   //to move the hook downward
    rescale the hook into y-axis about extAmount
    reposition the hook into y-axis about -1*extAmount
    count++
while count > 0 do
    rescale the hook into y-axis about -1*extAmount
    reposition the hook into y-axis about extAmount
    count--
```

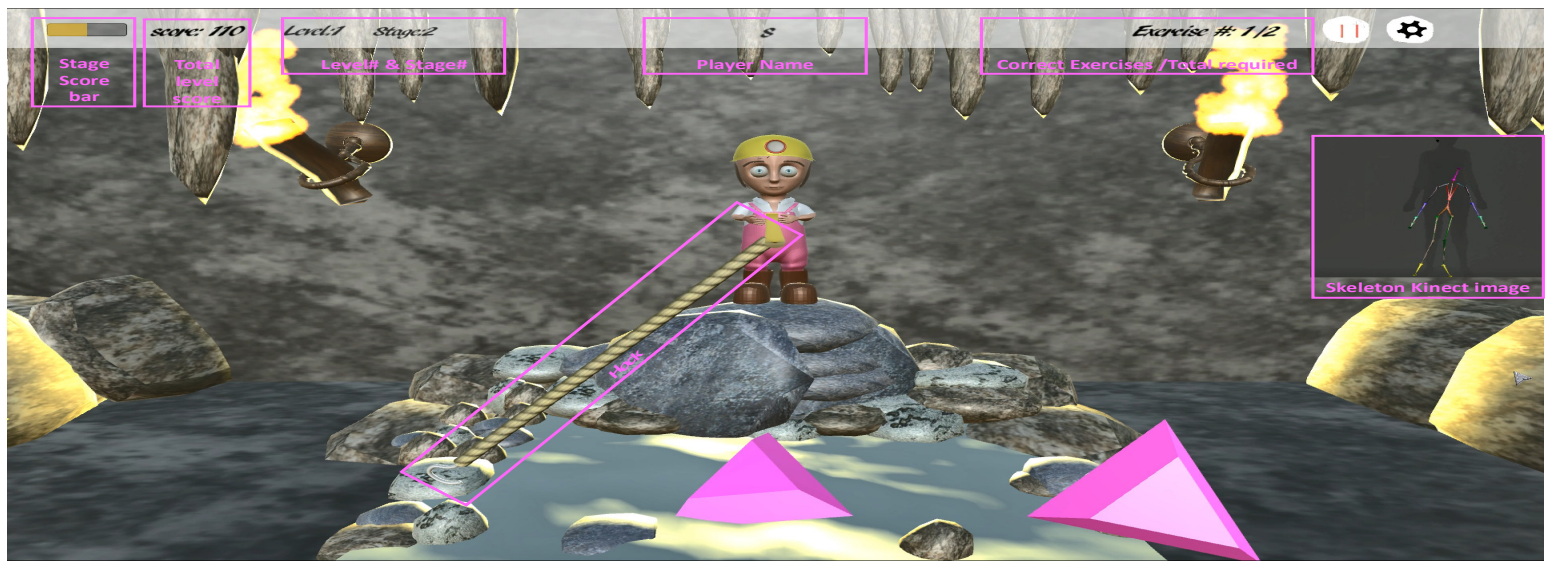

Figure 11. The main interface and the gamification feedback of the physical therapy section

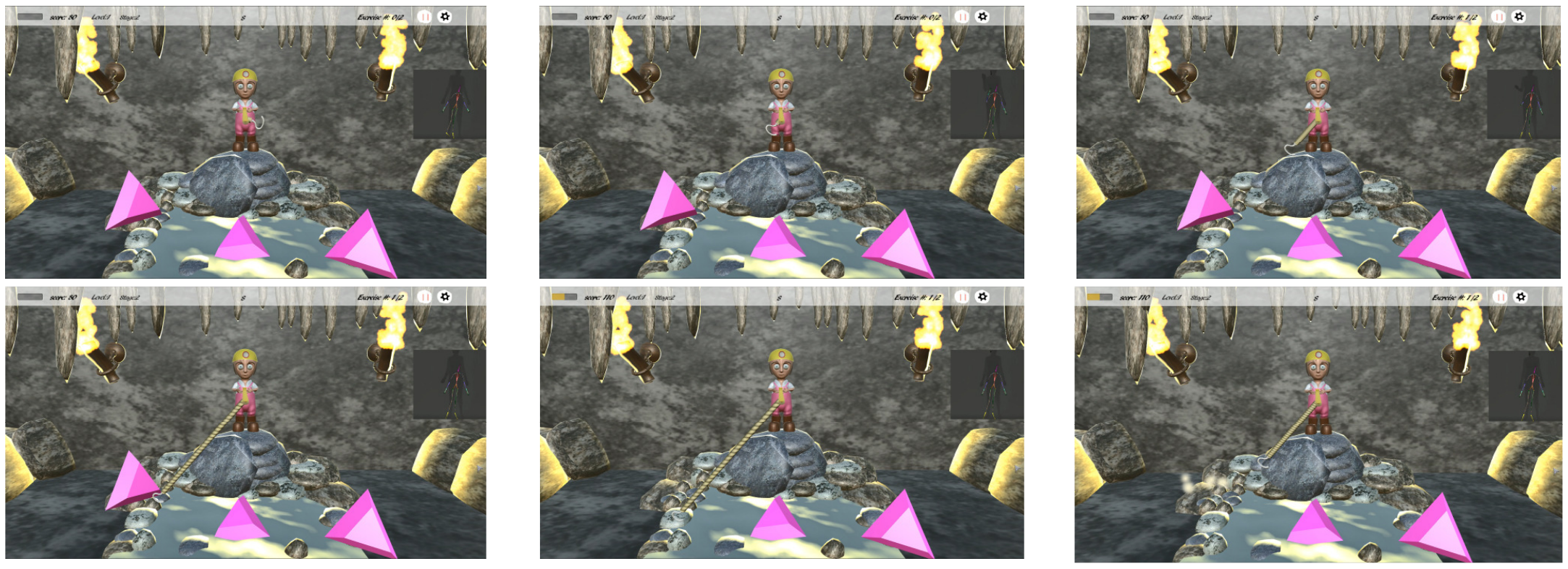

Figure 12. Hock movement for a complete correct elbow flexion/extension gesture

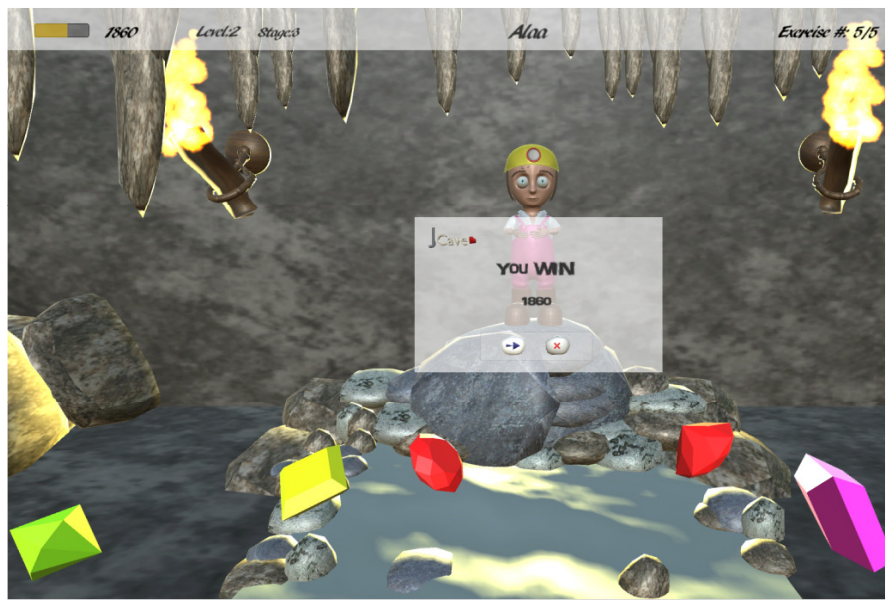

(a) Winning message

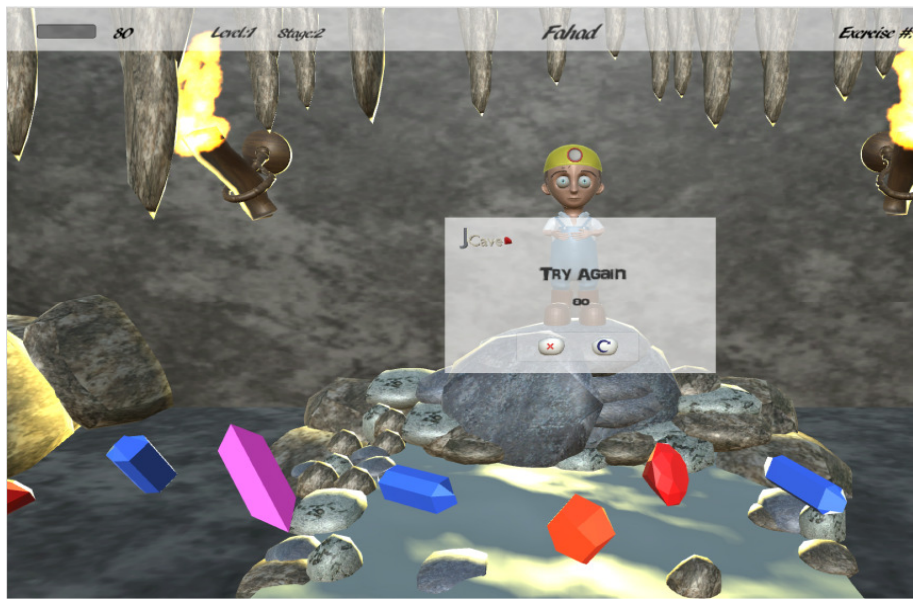

(b) Try again message

Figure 13. The main interface of the physical therapy section with feedback to the player

---

**Algorithm 3:** WinOrLose

---

**Input:** *collectedScore*, *N*, and *NofER*

// where *collectedScore* is the score the player earned in the sub-level/stage, *N* is the number of exercise repetitions, and *NofER* is the number of exercises the player completed.

**Output:** win or loss

---

**1:** **if** (*NofER* >= *N*) and (*collectedScore* >= *N**10) **then**
**2:** Player win
**3:** **else if** (*NofER* = *N**2) and (*collectedScore* < *N**10) **then**
**4:** Player lose

---

Table 5. Jewel types and the corresponding point values

| Jewel type | Jewel index | Size | | |
|---|---|---|---|---|
| | | 0 | 1 | 2 |
| 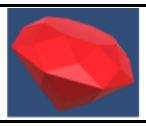 | 0 | 10 | 20 | 30 |
| 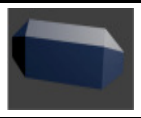 | 1 | 20 | 30 | 40 |
| 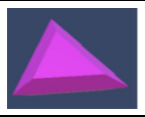 | 2 | 30 | 40 | 50 |
| 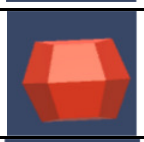 | 3 | 40 | 50 | 60 |
| 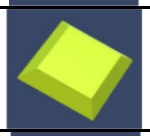 | 4 | 50 | 60 | 70 |
| 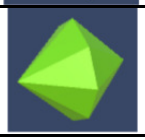 | 5 | 60 | 70 | 80 |

Table 6. Results of automatic gesture recognition testing

| Exercise | *h* = 148 cm, *d* = 1.5 m, and changing *t* | | *d* = 1.5 m and changing *h* at different *t* | | | *h* = 148 cm and changing *d* at different *t*. | | |
|---|---|---|---|---|---|---|---|---|
| | [a]*t* (sec) | Gesture recognition status | [b]*h* (cm) | *t* (sec) | Gesture recognition status | [c]*d* (m) | *t* (sec) | Gesture recognition status |
| Elbow flexion/ extension | 2.06 | Succeeded | 110 | 0.88 | Succeeded | 1 | - | Failed |
| | 6.35 | Succeeded | | | | 1.5 | 3.27 | Succeeded |
| | 4.81 | Succeeded | | | | 2 | 2.15 | Succeeded |
| | 3.27 | Succeeded | 155 | 4.91 | Succeeded | 2.5 | 1.9 | Succeeded |
| | 5.34 | Succeeded | | | | 3 | - | Failed |
| | 7.18 | Failed | | | | | | |
| Shoulder flexion | 1.98 | Succeeded | 110 | 1.06 | Succeeded | 1 | - | Failed |
| | 2.88 | Succeeded | | | | 1.5 | 2.33 | Succeeded |
| | 5.03 | Succeeded | | | | 2 | 4.02 | Succeeded |
| | 6.25 | Succeeded | 155 | 6.12 | Succeeded | 2.5 | 0.97 | Succeeded |
| | 4.57 | Succeeded | | | | 3 | - | Failed |
| | 7.32 | Failed | | | | | | |

[a]t: the time needed to perform the exercise in seconds
[b]h: height of the testers in centimeters
[c]d: the distance between the Kinect sensor and the testers in meters.

The JCave game is installed and runs on a laptop with Windows 10, which is a 64-bit operating system, with 8 GB of RAM, an Inter(R) Core™ i5-6200U CPU @ 2.30 GHz processor, and a GeForce 920M GPU. Four participants conducted the JCave test. Participants aged between 6 and 12 years. Figure 14 shows the JCave game testing setup. The participant in Figure 14 stands in front of the Kinect sensor at a distance of 1.4 m. The Kinect sensor is placed 0.6 m above the floor. The participant is 6 years of age, and his height is 120 cm. He created his profile and chose the elbow flexion/extension exercise for the left arm.

Table 7 shows the participants and the testing results in terms of the time taken to accomplish each task (T), the number of wrong selections (M), the number calls for help (H), the number of times a task was repeated after completing it successfully to measure the remember the process correctly (R), and the number of participants that felt frustrated (F) and confused (C).

Table 7 present the raw data that collected from the participants. We calculated the attributes for each task separately. Then we calculated the mean for all the tasks for each participant separately. The results show the participants completed all of the tasks successfully with excellent times. The overall JCave game results were good. Based on the participants' comments on the JCave game, the ability to explain how the exercises are performed while creating the profile and adding the ability to control the music and sound setting in the main menu were suggested.

To measure the level of satisfaction, all the participant rated the proposed game with no confusing and no frustrated which confirm the satisfaction of JCave game.

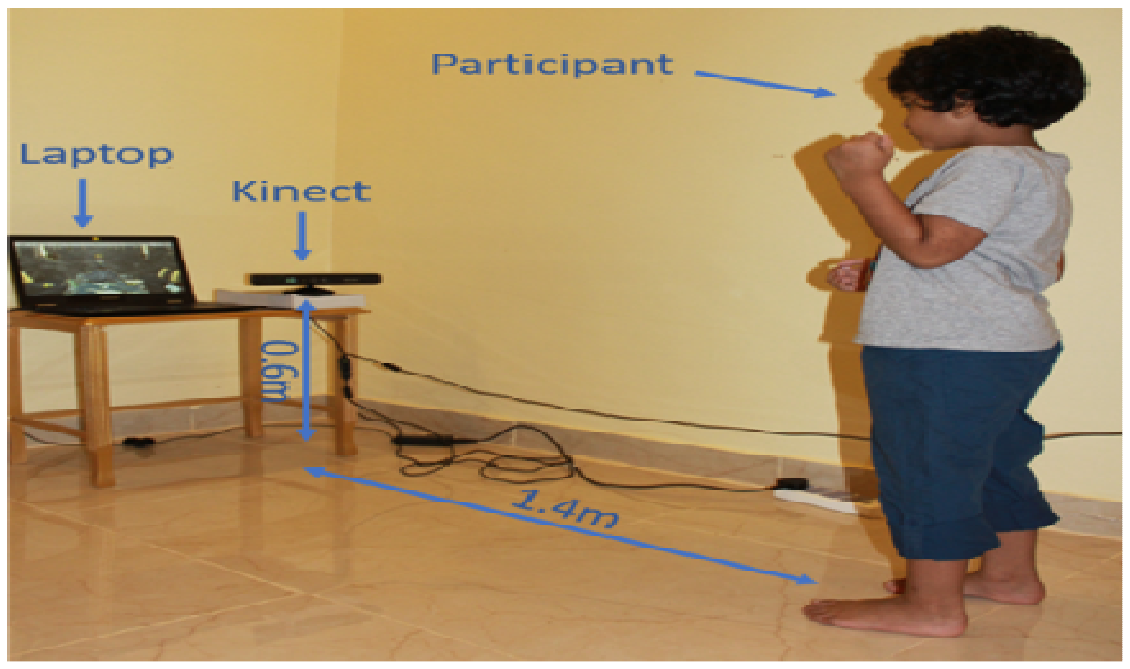

Figure 14. JCave game testing setup

Table 7. Result of the usability test

| Participant # | Measure** | *Task 1 | Task 2 | Task 3 | Task 4 | Task 5 | Task 6 | Task 7 | Task 8 | Task 9 | Task 10 | Task 11 | MeanTime for all the tasks per Participant |
|---|---|---|---|---|---|---|---|---|---|---|---|---|---|
| Participant #1 Age: 8 years Height: 122 cm | T (MM:SS) | 00:05 | 04:25 | 03:08 | 00:06 | 01:00 | 00:20 | 00:10 | 00:03 | 00:03 | 00:20 | 00:06 | 00.53 |
| | M | 0 | 0 | 0 | 0 | 0 | 0 | 0 | 0 | 0 | 0 | 0 | |
| | H | 0 | 0 | 0 | 0 | 0 | 0 | 0 | 0 | 0 | 0 | 0 | |
| | R | 0 | 0 | 0 | 0 | 0 | 0 | 0 | 0 | 0 | 0 | 0 | |
| | F | 0 | 0 | 0 | 0 | 0 | 0 | 0 | 0 | 0 | 0 | 0 | |
| | C | 0 | 2 | 0 | 0 | 0 | 0 | 0 | 0 | 0 | 0 | 0 | |
| Participant #2 Age: 6 years Height: 120 cm | T (MM:SS) | 00:08 | 01:07 | 00:14 | 00:04 | 01:00 | 00:24 | 00:03 | 00:05 | 00:15 | 00:20 | 00:09 | 00.21 |
| | M | 0 | 0 | 0 | 0 | 0 | 0 | 0 | 0 | 0 | 0 | 0 | |
| | H | 0 | 0 | 0 | 0 | 0 | 0 | 0 | 0 | 0 | 0 | 0 | |
| | R | 0 | 0 | 0 | 0 | 0 | 0 | 0 | 0 | 0 | 0 | 0 | |
| | F | 0 | 0 | 0 | 0 | 0 | 0 | 0 | 0 | 0 | 0 | 0 | |
| | C | 0 | 0 | 0 | 0 | 0 | 0 | 0 | 0 | 0 | 0 | 0 | |
| Participant #3 Age: 12 years Height: 145 cm | T (MM:SS) | 00:08 | 01:00 | 00:10 | 00:05 | 00:30 | 00:05 | 00:05 | 00:05 | 00:10 | 00:09 | 00:06 | 00.14 |
| | M | 0 | 0 | 0 | 0 | 0 | 0 | 0 | 0 | 0 | 0 | 0 | |
| | H | 0 | 0 | 0 | 0 | 0 | 0 | 0 | 0 | 0 | 0 | 0 | |
| | R | 0 | 0 | 0 | 0 | 0 | 0 | 0 | 0 | 0 | 0 | 0 | |
| | F | 0 | 0 | 0 | 0 | 0 | 0 | 0 | 0 | 0 | 0 | 0 | |
| | C | 0 | 0 | 0 | 0 | 0 | 0 | 0 | 0 | 0 | 0 | 0 | |
| Participant #4 Age: 12 years Height: 137 cm | T (MM:SS) | 00:06 | 00:50 | 00:11 | 00:11 | 00:10 | 00:03 | 00:03 | 00:05 | 00:07 | 00:07 | 00:04 | 00.11 |
| | M | 0 | 0 | 0 | 0 | 0 | 0 | 0 | 0 | 0 | 0 | 0 | |
| | H | 0 | 0 | 0 | 0 | 0 | 0 | 0 | 0 | 0 | 0 | 0 | |
| | R | 0 | 0 | 0 | 0 | 0 | 0 | 0 | 0 | 0 | 0 | 0 | |
| | F | 0 | 0 | 0 | 0 | 0 | 0 | 0 | 0 | 0 | 0 | 0 | |
| | C | 0 | 0 | 0 | 0 | 0 | 0 | 0 | 0 | 0 | 0 | 0 | |
| Mean Time per Task | | 00:07 | 01:50 | 00:56 | 00:06 | 00:40 | 00:13 | 00:05 | 00:04 | 00:09 | 00:14 | 00:06 | |

*Task 1: Start the game, Task 2: Create new profile, Task 3: Edit profile, Task 4: View help content, Task 5: Start playing, Task 6: Pause the game, Task 7: Resume the game, Task 8: Control the volume and music, Task 9: Exit play, Task 10: Delete profile, and Task 11: Exit from the game.
**T: the time taken to accomplish each task, M: the number of wrong selections, H: the number of calls for help, R: the number of times a task was repeated after completing it successfully, F: the number of times feeling frustrated, C: the number of times feeling confused.

## 7 Conclusion And Future Work

A 3D interactive game, called JCave, is designed for children from six to twelve years of age who need physical therapy for an upper limb. The game helps decrease the duration of the children's physical therapy rehabilitation program, motivates patients, and makes the home treatment exercises enjoyable and exciting. The AROM elbow flexion and extension and shoulder flexion exercise gestures are analyzed using the Xbox360 Kinect motion capture camera to capture the gestures and track the child. Based on these analyses an automatic gesture recognition is designed and implemented. The final JCave game consists of two main sections: the configuration/customization section and the physical therapy section. For the physical therapy section, the main interface and the gamification feedback design and implementation is presented for better user engagement. A usability test is applied to children between the ages of six and twelve, and based on the results, the game interface is easy to use. In the future, we will work on adding more exercises, adding more levels, saving the player progress, connecting the player progress with the physical therapy specialist, making an online version of the game, releasing the game to the public, and creating an Arabic version. Also, we plan in the future to have a control group of children who go through the rehabilitation process with/without using the game and report the progress of the rehabilitation process.

## ACKNOWLEDGEMENTS

The authors are grateful for all participants in the usability test.

## REFERENCES

[1] J. Stenros, "The Game Definition Game A Review," Games and Culture, vol. 12, no. 6, pp. 499-520, 2017. https://doi.org/10.1177/1555412016655679

[2] Tavinor, "Videogames as Art," in The Art of Videogames, Oxford, Wiley-Blackwell, 2009. doi: 10.1002/9781444310177.ch9

[3] Kontadakis, G., Chasiouras, D., Proimaki, D. et al., “Gamified platform for rehabilitation after total knee replacement surgery employing low cost and portable inertial measurement sensor node,” Multimed Tools Appl (2018). https://doi.org/10.1007/s11042-018-6572-6

[4] M. Hedström and I. Waernbaum, "Incidence of Fractures among Children and Adolescents in Rural and Urban Communities - Analysis Based on 9,965 Fracture Events," Injury Epidemiology, vol. 1:14, pp. 1-5, 2014. https://doi.org/10.1186/2197-1714-1-14

[5] L. v. Laer, Pediatric Fractures and Dislocations, Thieme, 1 edition (2004).

[6] M. E. Kho, A. Damluji, J. M. Zanni and D. M. Needham, "Feasibility and observed safety of interactive video games for physical rehabilitation in the intensive care unit: a case series," Journal of Critical Care, vol. 27, no. 2, pp. 219.e1-219.e6, 2012. https://doi.org/10.1016/j.jcrc.2011.08.017.

[7] M. Jr, BSPT and PTRP, "What is Physical Therapy ?," Physical Therapy notes.com, 27 July 2017. [Online]. Available: http://www.physicaltherapynotes.com/2010/08/physical-therapy.html. [Accessed January 2018].

[8] J. Han, L. Shao, D. Xu and J. Shotton, "Enhanced Computer Vision with Microsoft Kinect Sensor: A Review," vol. 43, no. 5, pp. 1318 - 1334, October 2013. doi:10.1109/tcyb.2013.2265378.

[9] "Active Range of Motion Exercises," drugs.com, 2017. [Online]. Available: https://www.drugs.com/cg/active-range-of-motion-exercises.html. [Accessed January 2018].

[10] M. Jr, BSPT and PTRP, "Range of Motion - Types of Range of Motion Exercises," Physical Therapy Notes.com, [Online]. Available: http://www.physicaltherapynotes.com/2010/11/range-of-motion-types-of-range-of.html. [Accessed 22 January 2018].

[11] "Elbow Fractures in Children," ortho info, October 2014. [Online]. Available: https://orthoinfo.aaos.org/en/diseases--conditions/elbow-fractures-in-children. [Accessed January 2018].

[12] H. S. Hasan and S. A. Kareem, "Human Computer Interaction for Vision Based Hand Gesture Recognition: A Survey," 2012 International Conference on Advanced Computer Science Applications and Technologies (ACSAT), Kuala Lumpur, 2012, pp. 55-60. doi: 10.1109/ACSAT.2012.37

[13] "ZED," stereolabs, [Online]. Available: https://www.stereolabs.com/. [Accessed 2017].

[14] J. W. Weingarten, G. Gruener and R. Siegwart, "A state-of-the-art 3D sensor for robot navigation," in 2004 IEEE/RSJ International Conference on Intelligent Robots and Systems (IROS), 2004. doi: 10.1109/IROS.2004.1389728

[15] N. Blanc, T. Oggier, G. Gruener, J. Weingarten, A. Codourey and P. Seitz, "Miniaturized smart cameras for 3D-imaging in real-time [mobile robot applications]," in Proceedings of IEEE Sensors, 2004, 2004. doi: 10.1109/ICSENS.2004.1426202

[16] R. Francese, I. Passero and G. Tortora, "Wiimote and Kinect: gestural user interfaces add a natural third dimension to HCI," in In Proceedings of the International Working Conference on Advanced Visual Interfaces (AVI '12), 2012. DOI=http://dx.doi.org/10.1145/2254556.2254580

[17] "optitrack," [Online]. Available: http://optitrack.com/. [Accessed 2017].

[18] "Kinect for Windows Sensor Components and Specifications," [Online]. Available: https://msdn.microsoft.com/en-us/library/jj131033.aspx. [Accessed 2017].

[19] Rollings and E. Adams, Andrew Rollings and Ernest Adams on Game Design, Indianapolis, IN: New Riders, 2003.

[20] K. K and E. SO, "Accuracy and Resolution of Kinect Depth Data for Indoor Mapping Applications.," Sensors, vol. 12, no. 2, pp. 1437-1454, 2012. doi:10.3390/s120201437

[21] Fossati, J. Gall, H. Grabner, X. Ren and K. Konolige, Eds., Consumer Depth Cameras for Computer Vision Research Topics and Applications, Springer-Verlag London, 2013. doi: 10.1007/978-1-4471-4640-7

[22] Shen Y, Ong SK, Nee AYC (2008), “An augmented reality system for hand movement rehabilitation,” In: Proceedings of the 2nd International Convention on Rehabilitation Engineering & Assistive Technology (pp. 189–192). Singapore Therapeutic, Assistive & Rehabilitative Technologies (START) Centre

[23] M. Field, D. Stirling, M. Ros, Z. Pan and F. Naghdy, "Inertial sensing for human motor control symmetry in injury rehabilitation," 2013 IEEE/ASME International Conference on Advanced Intelligent Mechatronics, Wollongong, NSW, 2013, pp. 1470-1475. doi: 10.1109/AIM.2013.6584302

[24] M. Pedraza-Hueso, S. Martín-Calzón, F. J. Díaz-Pernas and M. Martínez-Zarzuela, "Rehabilitation Using Kinect-based Games and Virtual Reality," Procedia Computer Science, vol. 75, pp. 161-168, 2015. https://doi.org/10.1016/j.procs.2015.12.233.

[25] "Mira," 2017. [Online]. Available: http://www.mirarehab.com/. [Accessed January 2018].

[26] M. A. Fraiwan, N. Khasawneh, A. Malkawi, M. Al-Jarrah, R. Alsa'di and S. Al-Momani, "Therapy central: On the development of computer games for physiotherapy," in 2013 9th International Conference on Innovations in Information Technology (IIT), Abu Dhabi, 2013. doi: 10.1109/Innovations.2013.6544388

[27] "Coordinate Spaces," microsoft, [Online]. Available: https://msdn.microsoft.com/en-us/library/hh973078.aspx. [Accessed 2017].

[28] "Tracking Users with Kinect Skeletal Tracking," [Online]. Available: https://msdn.microsoft.com/en-us/library/jj131025.aspx. [Accessed 2017].

[29] V. Pterneas, "IMPLEMENTING KINECT GESTURES," 27 January 2014. [Online]. Available: https://pterneas.com/2014/01/27/implementing-kinect-gestures/. [Accessed 2017].

[30] "Kinect setup on Xbox 360," Microsoft, [Online]. Available: https://support.xbox.com/en-GB/xbox-360/accessories/kinect-sensor-setup#475211aede4742d699510efef5b515f2. [Accessed 2017].